\documentclass[12pt,fleqn,aps]{revtex4}
\def\be{\begin{equation}}
\def\ee{\end{equation}}
\def\bea{\begin{eqnarray}}
\def\eea{\end{eqnarray}}
\def\ba{\begin{array}}
\def\ea{\end{array}}
\def\nn{\nonumber}
\def\p{\partial}
\def\ka{\kappa}
\def\ep{\epsilon}

\def\sda{\sin^2\theta}
\def\cda{\cos^2\theta}

\begin{document}
\pagestyle{myheadings}
\markboth{}{Scalar field equation in the Kerr-Sen geometry \hfill Wu $\&$ Cai~~}
\baselineskip 25pt

\title{\bf Massive complex scalar field in the Kerr-Sen geometry: \\
exact solution of wave equation and Hawking radiation}
\author{S.Q. Wu$^{1,2}$\footnote{E-mail: sqwu@ustc.edu.cn}
and X. Cai$^2$\footnote{E-mail: xcai@ccnu.edu.cn}}
\affiliation{\it\small $^1$Interdisciplinary Center for Theoretical
Study, $\&$ Department of Astronomy and Applied Physics, University
of Science and Technology, Hefei 230026, People's Republic of China \\
\it\small $^2$Institute of Particle Physics, Hua-Zhong Normal University,
Wuhan 430079, People's Republic of China}
\date{\today, revised version}

\begin{abstract}
\noindent
The separated radial part of a massive complex scalar wave equation in
the Kerr-Sen geometry is shown to satisfy the generalized spheroidal wave
equation which is, in fact, a confluent Heun equation up to a multiplier.
The Hawking evaporation of scalar particles in the Kerr-Sen black hole
background is investigated by the Damour-Ruffini-Sannan's method. It is
shown that quantum thermal effect of the Kerr-Sen black hole has the same
character as that of the Kerr-Newman black hole.


\end{abstract}
\maketitle

\newpage
\section*{I. INTRODUCTION}

In a recent paper,~$^1$ we have investigated exact solution of a massive complex
scalar field equation in the Kerr-Newman black hole background, and demonstrated
that both its radial part and its angular part can be transformed into the form
of a generalized spheroidal wave equation.~$^2$ Previous work on solution of a
massive scalar wave equation in the Kerr(-Newman) space-time had been completed
in Refs.~[3,4]. It is interesting to extend our analysis to solution of a scalar
wave equation in a Kerr-Sen black hole background.~$^5$ The Kerr-Sen solution
arising in the low energy effective string field theory is a rotating charged
black hole generated from the Kerr solution. The thermodynamic property of this
twisted Kerr black hole was discussed in Ref.~[6] by using separation of the
Hamilton-Jacobi equation of a test particle. The aim of this paper is to study
some exact solutions to a massive charged scalar wave equation and to find its
connection to the confluent Heun equation~$^7$ as well as to investigate quantum
thermal effect of scalar particles on the Kerr-Sen space-time.

The paper is organized as follows: In Sec. II, we separate a massive charged
scalar field equation on the Kerr-Sen black hole background into the radial
and angular parts. Sec. III is devoted to transforming the radial part into a
generalized spheroidal wave equation and to relating it to the confluent Heun
equation. Then, we investigate quantum thermal effect of scalar particles in
the Kerr-Sen space-time in Sec. IV. Finally, we summarize our discussions in
the conclusion section.

\section*{II. SEPARATING VARIABLES OF KLEIN-GORDON
EQUATION ON THE KERR-SEN BLACK HOLE BACKGROUND}

Constructed from the charge neutral rotating (Kerr) black hole solution, the
Kerr-Sen solution~$^5$ is an exact classical four dimensional black hole solution
in the low energy effective heterotic string field theory. In the Boyer-Lindquist
coordinates, the Kerr-Sen metric and the electromagnetic field vector potential
can be rewritten as:~$^6$
\bea
\qquad ds^2 &=& -\frac{\Delta}{\Sigma}\Big(dt -a\sda
d\varphi\Big)^2 +\frac{\sda}{\Sigma}\Big[adt
-(\Sigma +a^2\sda)d\varphi\Big]^2 +\Sigma
\Big(\frac{dr^2}{\Delta} +d\theta^2\Big) \nn \\
\qquad {\cal A} &=& \frac{-Qr}{\Sigma}(dt -a\sda d\varphi) \, ,
\eea
where $\Delta = r^2 +2(b -M)r +a^2 = (r -r_+)(r -r_-)$, $\Sigma = r^2 +2br
+a^2\cda$ and $r_{\pm} = M -b \pm \ep$ with $\ep = \sqrt{(M -b)^2 -a^2}$.

This metric describes a black hole carrying mass $M$, charge $Q$, angular
momentum $J = Ma$, and magnetic dipole moment $Qa$. The twist parameter
$b$ is related to the Sen's parameter $\alpha$ via $b = Q^2/2M = M \tanh^2
(\alpha/2)$. Because $M \geq b \geq 0$, $r = r_-$ is a new singularity in
the region $r \leq 0$, the event horizon of the Kerr-Sen black hole is
located at $r = r_+$. The area of the outer event horizon of the twisted
Kerr solution~$^5$ is given by $A_+ = 4\pi(r_+^2 +2br_+ +a^2)
= 8\pi Mr_+$.

We consider the solution of a massive charged test scalar field on the
Kerr-Sen black hole background (We use Planck unit system $G = \hbar = c
= k_B = 1$ throughout the paper). Because the Kerr-Sen metric (1) only
differs from the Kerr(-Newman) solution by the form of two functions
$\Delta$ and $\Sigma$, the Klein-Gordon field equation satisfied by
the complex scalar wave function $\Phi$ with mass $\mu$ and charge $q$
in such a space-time can be separated as $\Phi(t,r,\theta,\varphi) =
R(r)S_{m,0}^{\ell}(ka,\theta)e^{i(m\varphi -\omega t)}$, in which the
angular part $S_{m,0}^{\ell}(ka,\theta)$ is an ordinary spheroidal angular
wave function with spin-weight $s = 0$, while the radial part can be
given as follows:
\be
\qquad \p_r[\Delta\p_rR(r)] +\Big[\frac{(Ar -ma)^2}{\Delta} +k^2\Delta
+2Dr -\lambda \Big]R(r) = 0 \, , \label{rew}
\ee
here $\lambda$ is a separation constant, $A = 2M\omega -qQ$, $D = A\omega
-M{\mu}^2$, $k = \sqrt{{\omega}^2 -{\mu}^2}$ (We assume that $\omega > \mu$).
For later convenience, we also denote $\ep B = A(M-b) -ma$ and introduce
$W_{\pm} = (A \pm B)/2$.

With further substitution $R(r) = (r -r_+)^{i(A +B)/2}(r -r_-)^{i(A -B)/2}F(r)$,
we can transform Eq. (\ref{rew}) for $R(r)$ into a modified generalized
spheroidal wave equation with imaginary spin-weight $iA$ and boost-weight
$iB$ for $F(r)$:~$^1$
\be
\qquad \Delta\p_r^2F(r) +2[i\epsilon B +(1 +iA)(r -M +b)]\p_rF(r)
+[k^2\Delta +2Dr +iA -\lambda]F(r) = 0 \, .
\label{gswe}
\ee

Eq. (\ref{rew}) has two regular singular points $r = r_{\pm}$ with indices
$\pm iW_+$ and $\pm iW_-$, respectively, whereas Eq. (\ref{gswe}) has indices
$\rho_+ = 0$, $-2iW_+$ and $\rho_- = 0$, $-2iW_-$ at two singularities $r =
r_{\pm}$, respectively. The infinity is an irregular singularity of Eqs. (\ref{rew})
and (\ref{gswe}). Eq. (\ref{gswe}) has the same form as the radial part of the
massive complex scalar wave equation in the Kerr-Newman geometry~$^1$ with its
solution (when $\mu = 0$) named as the generalized spheroidal wave function.~$^2$
It is interesting to note that a special solution of function $F(r)$ satisfies
the Jacobi equation of imaginary index when $\omega = \pm \mu = qQ/M$ (namely,
$k =D = 0$).

\section*{III. GENERALIZED SPHEROIDAL
WAVE FUNCTION AND HEUN EQUATION}

In this section, we shall show that the generalized spheroidal wave equation
(\ref{gswe}) of imaginary number order is, in fact, a confluent form of Heun
equation.~$^7$ To this end, let us make a coordinate transformation
$r = M -b +\ep z$ and substitute $R(r) = (z -1)^{i(A +B)/2}(z +1)^{i(A -B)/2}
F(z)$ into Eq. (\ref{rew}), then we can reduce it to the following standard
forms of a generalized spheroidal wave equation:~$^{1,7}$
\bea
&&\qquad (z^2 -1)R^{\prime\prime}(z) +2zR^{\prime}(z) +\Big[(\ep k)^2(z^2 -1)
+2D\ep z \nn \\
&&\qquad\qquad\qquad +\frac{(Az +B)^2}{z^2 -1} +2D(M -b) -\lambda\Big]R(z) = 0 \, ,
\eea
and
\bea
&&\qquad (z^2 -1)F^{\prime\prime}(z) +2[iB +(1 +iA)z]F^{\prime}(z)
+[(\ep k)^2(z^2 -1) \nn \\
&&\qquad\qquad\qquad\qquad +2D\ep z +2D(M -b) +iA -\lambda]F(z) = 0 \, ,
\label{swe}
\eea
where a prime denote the derivative with respect to its argument.

The spin-weighted spheroidal wave function $F(z)$ is symmetric under the
reflect $k \rightarrow -k$. Letting $F(z) = e^{i\ep kz}G(z)$ without loss
of generality, we can transform Eq. (\ref{swe}) to
\bea
&&\qquad (z^2 -1)G^{\prime\prime}(z) +2[iB +(1 +iA)z
+i\ep k(z^2 -1)]G^{\prime}(z) \nn \\
&&\qquad +[2i\ep k(1 +iA -iD/k)z -2\ep kB +iA +2D(M -b)
-\lambda]G(z) = 0 \, .
\label{tswe}
\eea
By means of changing variable $z = 1 -2x$, we arrange the singularities
$r = r_+$ ($z = 1$) to $x = 0$ and $r = r_-$ ($z = -1$) to $x = 1$,
respectively, and reduce Eq. (\ref{tswe}) to a confluent form of Heun's
equation~$^{7,8}$
\be
\qquad G^{\prime\prime}(x) +\Big(\beta +\frac{\gamma}{x} +\frac{\delta}{x-1}
\Big)G^{\prime}(x) +\frac{\alpha\beta x -h}{x(x -1)}G(x) = 0 \, ,
\label{cH}
\ee
with $\gamma = 1 +2iW_+$, $\delta = 1 +2iW_-$, $\beta = 4i\ep k$,
$\alpha = -(1 +iA) +iD/k$, $h = \lambda -2i\ep k -iA +4\ep kW_+ -2Dr_+$.

This confluent Heun equation (\ref{cH}), with $h$ its accessory parameter,
has two regular singular points at $x = 0, 1$ with exponents ($0, 1 -\gamma$)
and ($0, 1 -\delta$),respectively, as well as an irregular singularity at
the infinity point. The power series solution in the vicinity of the point
$x = 0$ for Eq. (\ref{cH}) can be written as
\be
\qquad G(\alpha,\beta,\gamma,\delta,h;x)
= \sum\limits_{n = 0}^{\infty}g_n x^n \, ,
\ee
and the coefficient $g_n$ satisfies a three-term recurrence relation~$^{7,8}$
\bea
&&\qquad g_0 = 1 \, , \quad g_1 = -h/\gamma \, , \nn \\
&&\qquad (n +1)(n +\gamma)g_{n +1} -\beta(n -1+\alpha)g_{n -1}
= \big[n(n -1 -\beta +\gamma +\delta) -h\big]g_n  \, .
\eea
It is not difficult to deduce the exponent $1 -\gamma$ solution~$^8$ for $x = 0$
and obtain the power series solution in the vicinity of the point $x = 1$ by a
linear transformation interchanging the regular singular points $x = 0$ and $x = 1$:
$x \rightarrow 1 -x$. Expansion of solutions to the confluent Heun's equation in
terms of hypergeometric and confluent hypergeometric functions has been presented
in Refs.~[2,7]. The confluent Heun's functions can be normalized to constitute
a group of orthogonal complete functions.~$^7$ It should be noted that Heun's
confluent equation also admits quasi-polynomial solutions for particular values
of the parameters.~$^{7,8}$ It follows from the three-term recurrence relation
that $G(\alpha,\beta,\gamma,\delta,h;x)$ is a polynomial solution if
\bea
&&\qquad \alpha = -N \, , \quad\mbox{with integer}\quad N \geq 0 \, , \nn \\
&&\qquad g_{N +1}(h) = 0 \, ,
\eea
where $g_{N +1}$ being a polynomial of degree $N +1$ in $h$, that is, there are
$N +1$ eigenvalues $h_i$ for $h$ such that $g_{N +1}(h_i) \equiv 0$.

\section*{IV. HAWKING RADIATION OF SCALAR PARTICLES}

Now we investigate the Hawking evaporation~$^9$ of scalar particles in the
Kerr-Sen black hole by using the Damour-Ruffini-Sannan's (DRS) method.~$^{10}$
This approach only requires the existence of a future horizon and is completely
independent of any dynamical details of the process leading to the formation of
this horizon. The DRS method assumes analyticity properties of the wave function
in the complexified manifold.

In the following, we shall consider a wave outgoing from the event horizon
$r_+$ over interval $r_+ <r <\infty$. According to the DRS method, a correct
outgoing wave $\Phi^{\rm out} = \Phi^{\rm out}(t,r,\theta,\varphi)$ is an
adequate superposition of functions $\Phi_{r >r_+}^{\rm out}$ and $\Phi_{r
<r_+}^{\rm out}$:
\be
\qquad \Phi^{\rm out} = C\big[\eta(r -r_+)\Phi_{r >r_+}^{\rm out}
+\eta(r_+ -r)\Phi_{r <r_+}^{\rm out}e^{2\pi W_+}\big] \, ,
\ee
where $\eta$ is the conventional unit step function, $C$ is a
normalization factor.

In fact, components $\Phi_{r >r_+}^{\rm out}$ and $\Phi_{r <r_+}^{\rm out}$
have asymptotic behaviors:
\bea
&&\qquad \Phi_{r >r_+}^{\rm out} = \Phi_{r >r_+}^{\rm out}(t,r,\theta,\varphi)
\longrightarrow c_1(r -r_+)^{iW_+}S_{m,0}^{\ell}(ka,\theta)e^{i(m\varphi
-\omega t)} \, , \quad (r \rightarrow r_+) \\
&&\qquad \Phi_{r <r_+}^{\rm out} = \Phi_{r <r_+}^{\rm out}(t,r,\theta,\varphi)
\longrightarrow c_2(r -r_+)^{-iW_+}S_{m,0}^{\ell}(ka,\theta) e^{i(m\varphi
-\omega t)} \, , \quad (r \rightarrow r_+) \label{wf1}
\eea
when $r \rightarrow r_+$. Clearly, the outgoing wave $\Phi_{r >r_+}^{\rm
out}$ can't be directly extended from $r_+ <r <\infty$ to $r_- <r <r_+$,
but it can be analytically continued to an outgoing wave $\Phi_{r <r_+}^{\rm
out}$ that inside event horizon $r_+$ by the lower half complex $r$-plane
around unit circle $r = r_+ -i0$:
$$ r -r_+ \longrightarrow (r_+ -r)e^{-i\pi} \, .$$
By this analytical treatment, we have
\be
\qquad \Phi_{r <r_+}^{\rm out}\sim c_2(r -r_+)^{-iW_+}
S_{m,0}^{\ell}(ka,\theta)e^{i(m\varphi -\omega t)} \, . \label{wf2}
\ee

Eq. (\ref{wf1}) just takes one solution to the radial equation inside the
event horizon $r_+$, it has the same form of Eq. (\ref{wf2}) generated by
the analytical method. As $\Phi_{r >r_+}^{\rm out}$ differs $\Phi_{r
<r_+}^{\rm out}$ by a factor $(r -r_+)^{-2iW_+}$, then a difference
factor $e^{2\pi W_+}$ emerges due to the above analytical treatment.
Thus we can derive the relative scattering probability of the scalar
wave at the event horizon
\be
\qquad \Big|\frac{\Phi_{r >r_+}^{\rm out}}{\Phi_{r <r_+}^{\rm out}}\Big|^2
= e^{-4\pi W_+} \, ,
\ee
and obtain the thermal radiation spectrum with the Hawking temperature
$T = \ka/2\pi$.
\be
\qquad \langle {\cal N} \rangle = |C|^2 = \frac{1}{e^{4\pi W_+} -1} \, ,
\label{sptr}
\qquad W_+ = \frac{Ar_+ -ma}{2\ep}= \frac{\omega -m\Omega -q\Phi}{2\ka} \, ,
\ee
where the angular velocity at the horizon is $\Omega = a/2Mr_+$, the
electric potential is $\Phi = Q/2M = b/Q$, the surface gravity at the
pole is $\ka = (r_+ -M +b)/2Mr_+ = \ep/2Mr_+$.

The black body radiation spectrum (\ref{sptr}) demonstrates that the thermal
property of Kerr-Sen black hole is similar to that of Kerr-Newman black hole
though its geometry character is like that of the Kerr solution.~$^6$
Correspondingly, there exist four thermodynamical laws of the Kerr-Sen black
hole, similar to those of Kerr-Newman black hole thermodynamics.

\section*{V. CONCLUSION}

In this paper, we have shown that the separation of variables of the scalar
wave equation in the Kerr-Newman black hole background can apply completely
to the case of the twisted Kerr solution. The separated radial part can be
recast into the generalized spheroidal wave equation, which is, in fact, a
confluent form of Heun equation.

In addition, we find that the thermal property of the twisted Kerr black hole
resembles that of Kerr-Newman black hole though its geometry character is like
that of the Kerr solution. The Kerr-Sen solution shares similar four black hole
thermodynamical laws and quantum thermal effect as the Kerr-Newman space-time
does.

\acknowledgments

The authors thank the referee for his advice on improving this manuscript.
One of us (SQW) is very indebted to Dr. Jeff Zhao and Dr. C.B. Yang for their
helps in finding some useful references.

\end{document}